\documentclass[12pt,a4paper]{article}
\usepackage{amsmath}
\usepackage{amssymb}
\usepackage{graphicx}
\makeatletter
\def\mr@ignsp#1 {\ifx\:#1\@empty\else #1\expandafter\mr@ignsp\fi}%
\newcommand{\multiref}[1]{\begingroup%\let\protect\string%
\xdef\mr@no@sparg{\expandafter\mr@ignsp#1 \: }%
\def\mr@comma{}%
\@for\mr@refs:=\mr@no@sparg\do{\mr@comma\def\mr@comma{,}\ref{\mr@refs}}%
\endgroup} \makeatother 
    \def\be{\begin{eqnarray}}
    \def\ee{\end{eqnarray}}
    \def\no{\nonumber}
    \def\suml{\sum\limits}
    
    \def\intl{\int\limits}
    \def\intoinf {\int\limits_0^\infty}
    \def\intii{\int\limits_{-\infty}^{\infty}}
    
    \def\({\left(}
    \def\){\right)}
    \def\<{\left\langle\,}
    \def\>{\, \right\rangle}
    \def\[{\left[}
    \def\]{\right]}

    \def\la{\label}
    \def\a{\alpha}

    \def\g{\gamma}
    \def\G{\Gamma}
    \def\e{\epsilon}
    \def\z{\zeta}

    \def\s{\sigma}

    \def\p{\partial}
    \def\MC{{\mathbb{C}}}
    
    \def\MR{{\mathbb{R}}}
    \def\MZ{{\mathbb{Z}}}

    \def\hf{ {\textstyle{1\over 2}} }
     \def\qt{ {\textstyle{1\over 4}} }
   
   \def\SS{S}
   
\begin{document}
\thispagestyle{empty}
\begin{flushright}
{ }
{SPhT-t08/007}\\
\end{flushright}
\vspace{1cm}
\setcounter{footnote}{0}
\begin{center}
{\Large{
\bf 
Functional BES equation
}}

\vspace{20mm} { Ivan Kostov$^{\star }$, Didina Serban and Dmytro Volin
$^{\circ}$ } \\[7mm] {\it Service de Physique Th\'eorique, CNRS-URA
2306
\\
C.E.A.-Saclay \\
F-91191 Gif-sur-Yvette, France}\\[1mm]
 \end{center}

\vskip 9mm

\noindent{ We give a realization of the Beisert, Eden and Staudacher
equation for the planar ${\cal N}=4$ supersymetric gauge theory which
seems to be particularly useful to study the strong coupling limit.
We are using a linearized version of the BES equation as two coupled
equations involving an auxiliary density function.  We write these
equations in terms of the resolvents and we transform them into to a
system of functional, instead of integral, equations.  We solve the
functional equations perturbatively in the strong coupling limit and
reproduce the recursive solution obtained by Basso, Korchemsky and
Kota\'nski.  The coefficients of the strong coupling expansion are
fixed by the analyticity properties obeyed by the resolvents.}

\vfill

\noindent {\small $^\star$Associate member of the {\it Institute for
Nuclear Research and Nuclear Energy, \\ Bulgarian Academy of Sciences,
72 Tsarigradsko Chauss\'ee, 1784 Sofia, Bulgaria}}

 \vskip 4mm

\noindent {\small $^\circ$} {\small {On leave from \it Bogolyubov
Institute for Theoretical Physics, 14b Metrolohichna Str.  \\
Kyiv, 03143 Ukraine}.}

\section{Introduction and overview}

 One of the most exciting discoveries in the last few years is the
 integrability of the maximally supersymmetric Yang-Mills theory
 \cite{MZ02,BS03,BKS} and its relation to the superstring theory in
 $AdS^5\times S^5$ background \cite{MaldaAdS,GKP98,Witten98}.  An
 all-orders version of the Bethe Ansatz equations for AdS/CFT, valid
 asymptotically, was first formulated by Beisert, Eden and Staudacher
 \cite{BES}, based on previous work  \cite{ AFS,  HL, BHL}. 
  The proposal by Beisert, Eden and Staudacher
 solves the crossing symmetry conditions, formulated by Janik
 \cite{Janik}.  This proposal, interpolating all the way from strong
 to weak coupling, was intensively tested, the most sophisticated
 tests being performed on the large spin limit of the so-called
 twist-two, or cusp, anomalous dimension.  In the
 limit of large Lorentz spin $S$, this quantity scales logarithmically
 \cite{Korchemsky:1988si,  Korchemsky:1992xv,  
 Belitsky:2006en,ES,  Alday:2007mf}
 \be
 \Delta-S=f(g)\ln S+\ldots\;, 
 \ee 
 where
 $g$ is the coupling constant, related to the 't Hooft coupling
 constant $\lambda$ by
 \be g^2=\frac{\lambda}{16\pi^2}\;. 
 \ee

 The universal scaling function $f(g)$ was computed perturbatively in
 the gauge theory up to the fourth order.  The third order result was
 extracted \cite{KLOV} from a QCD computation by Moch, Vermaseren and
 Vogt \cite{Moch}.  The universal scaling function appears in the
 iterative structure of the multigluon amplitude \cite{Bern03} and it
 was computed to the third odrer in \cite{Bern05} and numerically to
 the fourth order, after an impressive effort \cite{Bern06,CSV}.  On
 the string side, the universal scaling function was also computed for
 the first three non-trivial orders
 \cite{Gubser:2002tv,Frolov:2002av,Roiban:2007jf,Roiban:2007dq}
\begin{equation}\label{strongBES}
f(g)=4\, g -\frac{3\,\log 2}{\pi} -\frac{\rm{K}}{4\,\pi^2} \frac{1}{g}
+\ldots\, ,
\end{equation}
where K$=\beta(2)$ is Catalan's constant.  It is remarkable that both
the weak coupling and the strong coupling results for the universal
scaling function can be reproduced from the conjectured Bethe ansatz
equations.  In this context, it is determined by the integral
equation, written down by Eden and Staudacher \cite{ES}
\begin{equation}
\label{BES}
\sigma(u)=\frac{1}{\pi}\int_{-\infty}^\infty{d v}\
\frac{\sigma(v)}{(u-v)^2+1}-\int_{-\infty}^\infty{d v}\;K(u,v)\ \(
\sigma(v)-\frac{1}{4\pi g^2}\)\;.
\end{equation}
With the integration kernel $K(u,v)$ determined in \cite{BES}, this
equation is known as the Beisert, Eden and Staudacher (BES) equation.
The universal scaling function is given by the integral of the density
\cite{Kotikov:2006ts}
\be
\la{deffg}
 f(g)=16g^2\int \sigma(u)\,du \, .
\ee

Although it is not, at least for the moment, possible to solve the
equation (\ref{BES}) in a closed form for arbitrary $g$, it is
relatively easy to extract from it the perturbative expansion at weak
coupling \cite{ES,BES}.  This coefficients of the perturbative
expansion agree with the field-theoretic results
\cite{KLOV,Bern05,Bern06,CSV}.

The strong coupling limit of the equation (\ref{BES}) proved to be
much more difficult to master analytically.  The first results were
obtained numerically \cite{Benna:2006nd} and they correspond to the
three coefficients in (\ref{strongBES}).  The first coefficient in
(\ref{strongBES}) was obtained analytically by various
methods \cite{Kotikov:2006ts,Alday:2007qf,Kostov:2007kx,Beccaria:2007tk},
while the second was obtained, although not from the BES equation, by
Casteill and Kristjansen \cite{Casteill:2007ct} and later by Belitsky
\cite{Belitsky:2007kf}.  The reason the strong coupling limit of the
BES equation is so difficult to take is that the expansion of the
scattering phase in powers of $1/g$ is not uniform in the rapidity
variable $u$.  There are three different regimes for $u$ which are to
be considered.  The first is the plane-wave limit \cite{BMN}, where
$|u/2g|\gtrsim 1$, or in terms of momenta $p\sim1/g$.  The second is
the so-called giant magnon regime \cite{Maldacena}, with
$|u/2g|\lesssim 1$ or $p\sim 1$.  The third regime was called
\cite{Maldacena:2006rv} the near-flat space regime, at it correspond
to to $u\pm2g\sim 1$, or to momenta of the order $p\sim 1/\sqrt{g}$.
As Maldacena and Swanson pointed out in \cite{Maldacena:2006rv}, in
this region the perturbative expansion of the dressing phase is
completely reorganized, compared to that in plane-wave and giant
magnon regimes.  This is the reason why the attempts to to solve the
BES equation order by order in $1/g$ failed beyond the leading
order\footnote{In the approaches
\cite{Casteill:2007ct,Belitsky:2007kf}, which in principle treat a
different limit than the BES equation, the contribution from the
near-flat space regime is suppressed from the beginning.  This
rapidity regime falls inside the gap of the density.} .

A big step forward was made recently by Basso, Korchemsky and
Kota\'nski \cite{Basso:2007wd}, who succeeded to give a procedure for
obtaining all orders in the strong coupling expansion recursively.
One of the important ingredients of their work was to linearize the
Bethe ansatz equations by transforming the BES equation into a set of
two equations, by exploiting the expression of the dressing kernel as
a convolution of two ``undressed" kernels \cite{BES}.  The result is a
set of two coupled integral equation, for the physical and an
auxiliary density.  The idea of linearization was first proposed by
Kotikov and Lipatov \cite{Kotikov:2006ts} and subsequently by Eden
\cite{Eden}.  Basso, Korchemsky and Kota\'nski \cite{Basso:2007wd}
used the Fourier representation of the BES equation, which is more
adapted for numerical analysis, as well as a number of numerically
inspired hypotheses.

The present work arose as an attempt to derive the results of
\cite{Basso:2007wd} by purely analytical consideration.  Our previous
work \cite{Kostov:2007kx} on the strong coupling limit of the BES
equation suggests that a natural way of solving this equation is to
reformulate it as a functional equation for the resolvent.  Here, we
will pursue this direction.  To use the resolvent to solve the Bethe
equations is by no means a new idea; in particular it was used in
\cite{Casteill:2007ct,Belitsky:2007kf}.  The novelty of our approach
is to replace the density in the BES equation (\ref{BES}) by the
resolvent and to exploit the analyticity property of the latter.
After writing the linearized BES equations in terms of the resolvent,
it is possible to transform the integral BES equations into a set of
functional equations for the physical and auxiliary resolvents.  This
procedure works for an arbitrary value of the coupling
constant.\footnote{The essential steps of this procedure are already
present in the paper \cite{Kotikov:2006ts} by Kotikov and Lipatov.  }
 
The question if the functional BES equations can be solved for any
value of the coupling constant is still open.  What is clearly
possible to do is to solve these equations at strong coupling in
perturbation theory, that is, by neglecting the non-perturbative
correction.  This is possible because, in the absence of
non-perturbative terms, the resolvents posses extra symmetry
properties.  The algorithm of solving the equations loosely follows
the one of Basso, Korchemsky and Kota\'nski \cite{Basso:2007wd}.

First, we find the general solution for the resolvents at strong
coupling in the giant magnons/plane-wave regimes.  In these regimes
the rescaled rapidity $u/2g$ is kept finite.  The functional equations
are linear and homogeneous, so that the general solution is a linear
combination of a countable set of particular solutions.  Each of these
functions have a non-integrable singularity at the points $u/2g =\pm
1$.  The indeterminacy in the coefficients of the linear combination
and the singularities of the individual solutions can be cured by
analyzing the solution in the near-flat space region, where the
perturbation series in $1/g$ is reorganized compared to the giant
magnon and plane-wave regimes.  In the near-flat space regime the
resolvents must have a series of integrable square root singularities
in the lower half-plane and must decrease as $1/u$ at infinity in the
upper half-plane.  Remarkably, the condition that the solutions in the
plane-wave/giant magnon and near-flat space regimes match analytically
allows to determine uniquely the resolvents in the strong coupling
limit up to non-perturbative corrections.  This matching condition is
equivalent to the quantization condition by Basso, Korchemsky and
Kota\'nski \cite{Basso:2007wd}.

 We formulate a recursive procedure for computing analytically the
 coefficients of the $1/g$ expansion of the density of Bethe roots.
 In the plane-wave regime, $|u/2g|>1$, the density has a fourth order
 branch point at $u/2g = \pm 1$ at any order in $1/g$.

The paper is organized as follows: in section 2 we derive the
linearized equations for the resolvents, in section 3 we transform the
integral equations into functional equations and we discuss the
analyticity properties of the resolvents and in section 4 we find the
perturbative solution by imposing the required analyticity properties
to the general solution.

 \subsection{Notations}

 The notations used in this paper are similar with those of our
 previous paper \cite{Kostov:2007kx}.  We will denote by $\e$ the
 inverse gauge coupling and will use a rescaled rapidity, more adapted
 to the strong coupling limit:
 \be\la{newu} \e\equiv\frac 1{4g}, \qquad u=\frac{u_{{\rm old}}}{2g}\,
 , \ee
 as well as the variable $x(u)$, related to $u$ by
 
 \be u(x)\equiv \frac 12\(x+\frac 1x\),\quad x(u)=u\(1+\sqrt{1-\frac
 1{u^2}}\)\;.  \ee
Note the branch cut of $x(u)$ for $u\in [-1,1]$.  In the intermediate
steps we will also use the notations
 \be x^{\pm }(u)
\equiv x(u \pm i\e) \, .
  \ee 
Sometimes it will be useful to switch to  the parametrization
which resolves the square root of $x(u)$ and which is the hyperbolic
limit of the elliptic parametrization in \cite{Kostov:2007kx}:
\be
  u = \coth s\, , \ \ \ \ x (s)=
\coth {s\over 2}, \qquad  {u+1\over u-1}= e^{2s}\, .
\ee
The BES equation is formulated for the density function $\s(u)$ in the
limit of large spin $S\gg1$, related to the distribution $\rho(u)$ of
Bethe roots by
\be \la{defrho} \rho(u) = \( {2\e\over\pi} - \s(u) \) {\log S} + {\cal
O}(S^0)\,,\qquad u\ll S/2g .  \ee

\section{Linearized BES equations for the resolvent}

In this section, we reformulate the Beisert, Eden and Staudacher
equation \cite{ES,BES}, which determines the density of Bethe roots
corresponding to the twist-two operator, as a set of two equations for
the physical resolvent and an auxiliary resolvent.  This is
essentially the program which was carried out by Kotikov and Lipatov
\cite{Kotikov:2006ts} and by Eden \cite{Eden}, although they have not
explicitly identified the inverse Fourier transform of the density on
the positive half-axis as the resolvent.

 \subsection{The resolvent}

Let us consider the density $\sigma(u)$, supported on the real axis,
as well as its Fourier transform\footnote{Our definition of
$\sigma(t)$ is slightly different of that of \cite{ES} and the
subsequent references.} $\sigma(t)$
\begin{equation}
 \sigma(t)=\int_{-\infty}^\infty du\; e^{itu}\sigma(u)\;.
\end{equation}
The resolvent is defined, as usually, by
\be R_{\textrm{phys}}(u)=\intii dv \;\frac
{\s(v)}{u-v}\, . 
\ee
and it can be seen as the inverse Fourier transform of the density
$\sigma(t)$ on the positive half-axis.  Let us assume that $u$ is in
the upper half plane.  Then we can write
\be R_{\textrm{phys}}(u)&=&-i\int _{-\infty}^\infty dv \int_0^\infty
dt\; e^{it(u-v)}\sigma(v) \\ &=& -i\int_0^\infty dt\;
e^{itu}\sigma(-t)\, .
\ee
The density is given by the discontinuity of the resolvent across the
real axis:
\be\la{density} \s(u)= \frac 1{2\pi
i}\[R_{\textrm{phys}}(u-i0)-R_{\textrm{phys}}(u+i0)\] \ee
Since the density is supported by the whole real axis, the resolvent
is given by two different analytic functions in the upper and in the
lower half-planes.  Due to the symmetry $\s(-u)=\s(u)$ we have the
relation
\be R_{\textrm{phys}}(-u)=-R_{\textrm{phys}}(u)
\ee
Therefore, the symmetry determines the resolvent in the lower half
plane once the resolvent in the upper half plane is known.  The
behavior of the resolvent at infinity is related to the universal
scaling function
\be\la{largeuR} R_{\textrm{phys}}(u)\sim \frac{1}{u}{\intii dv\, \s
(v)}=\frac{1}{u}\;\frac{f(g)}{16g^2}\;.  
\ee

 \subsection{Linearizing the BES equation }

 The linearization of the BES equation is best performed on the
 Fourier transformed form, although the manipulations can be done
 abstractly without reference to a particular representation.  In our
 notations, the Fourier transformed BES equation reads
\be 
\label{fourier}
(1-e^{-2\e t})\sigma(t)=-\int_0^\infty \frac{dt'
}{2\pi}\(K(t,t')+K_d(t,t')\)(\sigma(t')-\sigma_0(t'))\;.  \ee
where $\sigma_0(t)=4\e \delta(t)$.
Here we use the conventions
\be K(t,t')=2\pi
te^{-(t+t')\e}\sum_{n>0}2n\frac{J_n(t)J_n(t')}{tt'}\equiv
K_+(t,t')+K_-(t,t')\;,
\ee
where $K_+$ and $K_-$ contain the expansion on {\it odd} and {\it
even} order Bessel functions respectively, and the dressing kernel is
given by the ``magic formula'' \cite{BES}
\be K_d(t,t')=2\int_0^\infty \frac {dt''}{2\pi}
K_-(t,t'')\frac{1}{1-e^{-2\e t''}}K_+(t'',t')\;.  
\ee
Two representations of the dressing kernel in the rapidity space were given
 in \cite{Kostov:2007kx} and \cite{Dorey:2007xn}.
It is convenient to introduce an operator $\SS$, diagonal in Fourier
representation:
    \be 
    \SS(t)=\frac 1{1-e^{-2\e t}} \, .
    \ee
In the rapidity space  
    \be \SS^{-1} = 1- D\, , \quad {\rm where} \quad \la{defD} D
    f(u)=f(u+2i\e)\;.  \ee
The equation (\ref{fourier}) can be written symbolically as
\be 
-2\sigma=[(1+2\SS K_-)(1+2\SS K_+)-1](\sigma- \sigma_0)\, .
\ee
Now it is possible to transform the BES equation into a pair of
equations with the ``main'' kernels $K_\pm$ appearing linearly.  This
can be done at the expense of introducing an auxiliary density $\tau$
defined by
\be 
\tau+\sigma_0\equiv-(1+2\SS K_+)(\sigma- \sigma_0)\, .
\ee
The linear system of equations obeyed by the physical and auxiliary
density $\sigma$ and $\tau$ is simply
\be 
\label{decoup}
\tau+\sigma&=&-2\SS K_+(\sigma-\sigma_0)\, \\
\tau-\sigma&=&-2\SS K_-(\tau+\sigma_0)\;, \nonumber
\ee
or, in Fourier representation,
\be 
\label{decfour}
(1-e^{-2\e t})(\tau(t)+\sigma(t))&=&-2\int_0^\infty \frac{dt'
}{2\pi}K_+(t,t')\, (\sigma-\sigma_0)(t')\, , \no \\
(1-e^{-2\e t})(\tau(t)-\sigma(t))&=&-2\int_0^\infty \frac{dt'
}{2\pi}K_-(t,t')\, \tau(t')\;.  \ee
In the last line we have used  that $K_-(t,0)=0$.

\subsection{ Holomorphic  BES kernels}

The next step is to transform back the equations (\ref{decfour}) in
the rapidity space.  As we mentioned above, the inverse half-space
Fourier transform of the density $\s(t)$ gives the resolvent $R_{\rm
phys}(u)$.  The latter defines a pair of functions $R^{\rm up}(u)$ and
$R^{\rm down}(u)= -R^{\rm up}(-u)$, analytic respectively in the upper
and lower rapidity half-planes.  Because of the symmetry property, we
are going in the following to concentrate exclusively on $R^{\rm
up}(u)$.  It is this function, together with its analytical
continuation beyond the real axis, which will be denoted in the
following by $R_{\rm phys}(u)$.

Assume that $\Im u>0$ and perform the half-space inverse Fourier
transformation to the rapidity plane,
    \be \intoinf \frac{dt}{2\pi} e^{itu}\intoinf \frac{dt'}{2\pi}
    K(t,t')f(t')=\intii dv \intoinf \frac{dt}{2\pi} e^{itu}\intoinf
    \frac{dt}{2\pi} e^{-it'v}K(t,t') \ \intoinf \frac{dt''}{2\pi}
    e^{it''v}f(t'') \, .  \ee
Therefore, since we intend to work only with functions defined in the
 upper half plane, we can retain only half of the original kernel in rapidity 
 space, namely
    \be K^\e(u,v)=\intoinf \frac{dt}{2\pi} \intoinf \frac{dt'}{2\pi}
    e^{itu-it'v}K(t,t')\; .\ee
Here we use the superscript $\e$ for the kernel in order to indicate
that it depends on the coupling constant $g=1/4\e$.  Explicitly the
``holomorphic'' part of the odd and the even kernels reads
\be K^\e_-(u,v)&=&-\frac{1}{2\pi i}
\frac{d}{du}\Big[\ln\(1-\frac{1}{x^+y^-}\)+\ln\(1+\frac{1}{x^+y^-}\)\Big]\no
\\
\label{kernels}
K^\e_+(u,v)&=&-\frac{1}{2\pi i}
\frac{d}{du}\Big[\ln\(1-\frac{1}{x^+y^-}\)-\ln\(1+\frac{1}{x^+y^-}\)\Big]\;
.  \ee
The dependence on $\e$ in (\ref{kernels}) comes only through the
shifts $x^\pm=x(u\pm i \e)$ and it will be removed by change of
variable and shift of the integration contour.  The $\e\to 0$ limit of
the kernels (\ref{kernels}) will be denoted without superscript
    \be K_-(u,v)&=&\frac 1{2\pi i}\frac 2{1-x^2}\(\frac 1{y-\frac
    1x}-\frac 1{y+\frac 1x}\)\no\\
    \label{kernelss}
     K_+(u,v)&=&\frac 1{2\pi i}\frac 2{1-x^2}\(\frac 1{y-\frac
     1x}+\frac 1{y+\frac 1x}\)\, .
 \ee When using these equations, one has to remember that $\Im u>0$
 and $\Im v<0$, or
    \be 
    x= x(u+i 0),\quad y = y(v- i 0)=1/y(v+i0)\, . 
    \ee

\subsection{BES equations for the resolvents }

 For later convenience, we introduce the shifted resolvents
    \be R(u)&=&-i\intl_0^\infty dt\; e^{iut}\, e^{\e
    t}\;\s(t)=R_{\textrm{phys}}(u-i\e)\no\\
    H(u)&=&-i\intl_0^\infty dt\; e^{iut}\, \e^{\e
    t}\;\tau(t)\, .  \ee
 This definition is valid for $\Im[u]>0$ and the first singularity for
 $R(u)$ and $H(u)$ is situated on the real axis.  $R$ and $H$ can be
 analytically continued to $\Im[u]<0$.  We will introduce another pair
 of functions by
 \be
 R_\pm(u)
   = \hf [R(u)\pm H(u)]  
   \ee
as well as the related functions $r_\pm(u)$
\be
  % R_\pm=\hf (R\pm H) = S r_\pm\;,\quad
  r_\pm(u) = R_\pm(u)-R_\pm (u+2i\e)\, .
 \label{defrpm}
  \ee

Now we can take the inverse Fourier transform of the equations
(\ref{decfour}) and make the shifts $u\to u-i\e$ and $v\to v+i\e$, in
order to get rid of the $\e$-dependence in the kernels $K^\e_\pm
(u,v)$.  We obtain the equations:\footnote{This way of rewriting the
BES equation have been first proposed by Kotikov and Lipatov in
\cite{Kotikov:2006ts}.  }
 \be\label{eq:theequationpm} 
 R_+(u)-R_+(u+2i\e)& = &\frac
 {4i\e}{x^2-1}-\int dv \; K_+(u,v)\, [R_+(v+2i\e)+R_-(v+2i\e)]\no\\
  R_-(u)-R_-(u+2i\e)&=&\int dv\; K_-(u,v)\,
  [R_+(v+2i\e)-R_-(v+2i\e)]\, .  
  \ee
       %
%  \be\label{eq:theequation} R(u)-R(u+i\e)+H(u)-H(u+i\e) &=&\frac
%  {4i\e}{x^2-1}-2\int \limits_{-\infty}^\infty dv
%  K_+(u,v)R(v+i\e)\no\\ {R(u)-R(u+i\e)-H(u)+H(u+i\e) }&=&2\int
%  \limits_{-\infty}^\infty dv K_-(u,v)H(v+i\e)\;.  \ee
%
%\be\label{eq:theequation} r_+(u) &=&\frac {2i\e}{x^2-1}-\int
%\limits_{-\infty}^\infty dv\; K_+(u,v)\;R(v+i\e)\no\\ {r_-(u)}&=&\int
%\limits_{-\infty}^\infty dv \;K_-(u,v)\;H(v+i\e)\;.  
%\ee
%
 After   the change of variables the integration contour  for $v$ goes 
 along the shifted real axis   $\MR -i\e$, but it can be placed 
anywhere between the branch cut
$[-1,1]$ of the kernels $K_\pm$ and the branch cut $[-1-2i\e, 1-2i\e]$
of the resolvent $R(u)$.  We will assume that the contour is just below 
the real axis.  The variable $u$ originally lies in $\Im u>0$, but we can
analytically continue it to the whole rapidity plane, using that the 
integration kernels are holomorphic.

\section{Functional equation}

The kernels (\ref{kernelss}) look almost like Cauchy kernels, if not
for the branch cut of the variables $x(u)$ and $y(v)$.  This suggests
that we may simplify the BES equation further.  This can be done
provided that we know the analytical properties of the functions the
kernels act on.  In this section, we derive the analytic properties of
the resolvents $R_\pm(u)$ and of the functions $r_\pm(u)$ and
translate the action of the kernel in terms of an integral on the
interval $[-1,1]$.  This transformation will allow to transform the
integral equations into a functional equation.

\subsection{Analytic properties of the resolvents}

We start with the linearized BES equations
  \be
     r_+(u)&=&\frac {4i\e}{x^2-1}-\!\!\int\limits_{\MR-i0}
     \!\! dv K_+(u,v)\, [R_+(v+2i\e)+R_-(v+2i\e)]\no\\
     r_-(u)&=& \int\limits_{\MR-i0}\!\!  dv K_-(u,v)\,
     [R_+(v+2i\e)-R_-(v+2i\e)] \ee
    where
    \be 
 K_\pm(u,v)&=&\frac 1{2\pi i}\frac 2{1-x^2}\(\frac 1{y-\frac
    1x}\pm\frac 1{y+\frac 1x}\) 
    \ee
with $x= x(u)$ and $y = x(v)$.  The variables $u$ and $v$ belong to
the physical sheet, which means that $|x|>1$ and $|y|>1$.  Since the
kernel becomes singular only for $u$ in the interval $[-1,1]$, and
there is no other singularity when $u$ and $v$ are on the physical
sheet, the functions $r_\pm(u)$ are analytic in $\MC\backslash
[-1,1]$.  We deduce that the resolvents
\be
  R_\pm(u)=\suml_{n=0}^\infty r_\pm(u+2in\e)\;,
\ee
have a semi-infinite set of equidistant cuts as shown in Fig. 1.  
From the explicit form of the kernels it follows
that
  \be r_+(u)\propto \frac 1{u^2}\; ,\; \qquad r_-(u)\propto \frac
  1{u^3}\; \; \qquad (u\rightarrow\infty) \, .
\label{asyRr}
\ee
 The large $u$ behavior of the resolvents is  
  $R_+(u) \propto 1/u$ and $R_-(u)\propto 1/u^2$ .

\vskip 1.5cm

%%%%%%%%%%Fig.1 %%%%%%%%%%%%%%%%%%%% 
 \begin{center}
\includegraphics[scale=0.6]{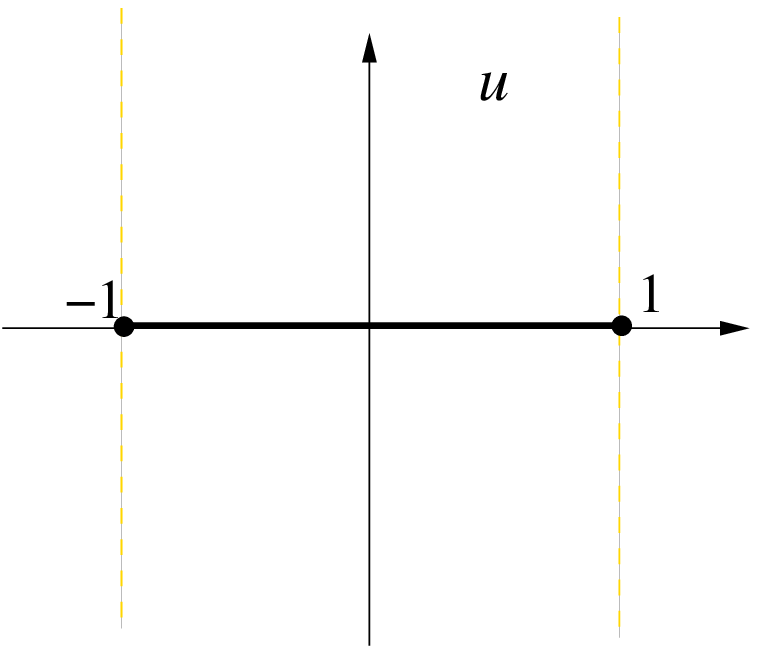}
\hskip 2cm
\includegraphics[scale=0.6]{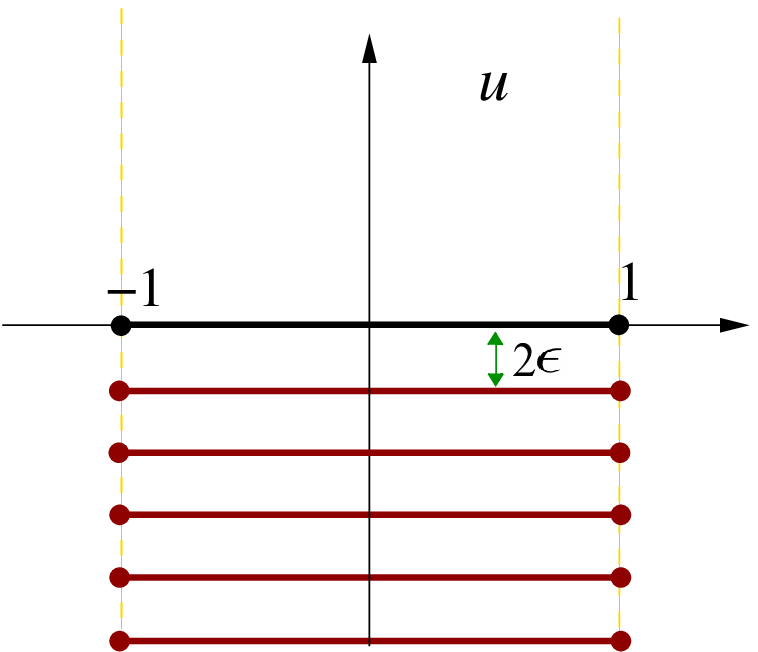}
\end{center}
\bigskip
\begin{center}
{\small Fig.  1:  Left: Physical sheet for $r_\pm(u)$.
Right: Physical sheet for $R_\pm(u)$. }
\end{center}
%%%%%%%%%%%%%%%%%%%%%%%%%%%%%%%%%   
  \vskip 1cm

\subsection{Analytic properties of the kernels.}

Since the resolvents have definite analyticity properties, we can
considerably simplify the action of the integration kernels.  Consider
the integral
\be
\intl_{-\infty-i0}^{+\infty-i0} dv\; K_\pm(u,v) F(v)\,, 
\ee
where $F(v)$ stands either for $R_+(v+2i\e)$ or for $R_-(v+2i\e)$.
The function $F(u)$ is analytic in the half-plane $\Im u>-\e$.  We
will actually need a weaker assumption, namely that the function
$F(u)$ is analytic in the upper half plane $\Im u\ge 0$ with the real
axes included.
  
We would like to place the integration contour above the real axis,
since in the upper half-plane both $K_\pm$ and $F$ are analytic.
Using the properties \be x(v-i0)&=& 1/x(v+i0) \;,\; u\in [-1,1]\no\\
  x(v-i0)&=&x(v+i0)\;,\;u\in \MR\backslash [-1,1]\,
\ee
we can place the integration contour above the real axis at the price
of changing the form of the kernel in the interval $[-1,1]$.  Next,
since in the upper half-plane the function $F$ is analytic, we can
deform the rest of the contour to to a contour that goes along the
interval $[-1,1]$ in the opposite direction.  Adding up the two
contributions we evaluate the integral as
  \be \la{eq:kernewaction} K_\pm F (u)&=&\frac
  2{1-x^2}\intl_{-1+i0}^{1+i0}\frac {dv}{2\pi i} F(v)\(\frac
  {-yx}{y-x}\pm\frac {yx}{y+x}-\frac 1{y-\frac 1x}\mp\frac 1{y+\frac
  1x}\)\no\\&=& \intl_{-1+i0}^{1+i0}\frac {dv}{2\pi i} F(v )\frac
  {y-\frac 1y}{x-\frac 1x}\(\frac 1{v-u}\mp\frac 1{v+u}\)\, .  \ee
In the original equation (\ref{eq:theequationpm}), the integration
contour is pinched between two cuts distanced by $\e$ and the limit
$\e\to 0$ is not well defined.  On the contary, the form
(\ref{eq:kernewaction}) of the integration kernel has a smooth $\e\to
0$ limit:

\be\label{eq:fdefinition}
  K_\pm F
%  &\equiv& \intl_{-1+i0}^{1+i0}\frac {dv}{2\pi i} F[v]\ \frac
%  {y-\frac 1y}{x-\frac 1x}\(\frac 1{v-u}\mp\frac 1{v+u}\) \no\\
  &=&\intl_{-1}^{1}\frac {dv}{2\pi   } \sqrt{\frac {  1- v^2} { u^2-1
  }}\, \frac { F(v+i0)\pm F(-v+i0)}{v-u } \, . 
  \ee

\subsubsection{Projective properties of the kernels $K_\pm$}

Via the transformation $v\rightarrow -v$ the integral in
(\ref{eq:fdefinition}) can be written as an integral over a closed
contour around the segment $[-1,1]$:
  \be K_\pm F=\oint \frac {dv}{2\pi i} \ \tilde F(v) \, \sqrt{\frac {
  v^2-1} { u^2-1 }}\, \frac 1{v-u} \ee with \be \tilde F(u)\, =\,
  \begin{cases}
     \ \ \ F(u)  & \text{  if  $\Im u>0 $  }, \\
      \pm F(-u) & \text{ if  $\Im u<0$}\, .
\end{cases}
\ee

Denote by ${\cal L^\pm}$ the linear space of even/odd functions,
analytic outside the interval $[-1,1]$ and decreasing at infinity
faster than $1/u$ at infinity.  For any $f_\pm\in {\cal L^\pm} $ , the
kernel $K_\pm$ acts as the identity operator:
\be\la{Proj} K_{\pm} f_\pm=f_\pm ,\quad f_\pm \in \cal L^\pm \, .  \ee
In particular, 
\be\la{Krr} K_+r_+= r_+, \quad K_- r_-=r_-\, .  
\ee
Now consider the result of the action of the kernel $K_\pm$ on an
arbitrary function $F(u)$.  Since the kernel $K_\pm(u,v)$ is even/odd
as a function of $u$, is analytic outside the interval $[-1,1]$, and
decreases as $1/u^2$ at infinity, the resulting function $F_\pm=K_\pm
F$ belongs to ${\cal L}^\pm$.  As a consequence, $K_\pm^2 F = K_\pm
F_\pm = F_\pm = K_\pm F$.  We conclude that the kernel $K_\pm$ is
idempotent:
\be K_\pm^2=K_\pm\, .  \ee

An example of a function that does not belong to ${\cal L^\pm}$ is the
constant function.  The action of $K_\pm$, evaluated by expanding the
contour to infinity, is
 \be
 K_+ \cdot 1= -{1/x \over \sqrt{u^2-1} } =
 \frac 2{1-x^2} ,
 \quad K_- \cdot 1 = 0 \, .
 \la{Kone}
 \ee

\subsection{The BKK transformation}

The linearized BES equations (\ref{eq:theequationpm}) considerably
simplify when written in terms of the functions $\G_{+} $ and $\G_{-} $
defined by\footnote{This transformation corresponds to the one given
by eq.  (6) in \cite{Basso:2007wd}.}
 \be\la{defGpm}
  \G_{+} (u)+\G_{-} (u)&\equiv&R_+(u)+R_-(u+2i\e)+2i\e\no\\
  \G_{+} (u)-\G_{-} (u)&\equiv&R_-(u)-R_+(u+2i\e)+2i\e\, .
\ee
Indeed, with the help of the identities
 (\ref{Kone}) and  (\ref{Krr}) we write
 the linearized BES equations (\ref{eq:theequationpm})
 as
\be\la{KGG}
K_+( \G_{+}  + \G_{-}  )
%=\hf (1- K_+) r_+
=0,
\\
K_-(\G_{+}  -\G_{-}  )
%= -\hf (1- K_-) r_-
=0\, .
\ee
Therefore the solution of the BES equation is a linear combination of
the zero modes of the operators $K_\pm$.

%\subsubsection{Zero modes of $K_\pm$}

We have seen that the action of the kernels $K_\pm$ on the functions
analytic in the upper half plane is given by (\ref{eq:fdefinition}).
As a consequence, the necessary and sufficient condition that the
function $F$ is annihilated by $K_\pm$ is
\be\label{eq:evenoddsolution}
  F(u+i0)\pm F(-u+i0)=0\;,\; u\in [-1,1]\, .
\la{FeF}
\ee
Note that the condition (\ref{eq:evenoddsolution}) does not imply the
function $F$ is odd or even.  It can be written in terms of the
variable $x$ as
\be\la{FEX} F(x)=\mp F(-1/x) \, .\ee For example the first equation
(\ref{Kone}) means that $f_0(x) = {1+x^2\over 1-x^2}$ is a zero mode
of $K_+$.  This function satisfies $f_0(-1/x)= - f_0(x)$.

\subsection{From integral to functional
equations}\label{subsec:funceq}

Now we can reformulate the homogeneous integral equations (\ref{KGG})
as a pair of functional equations for $\G_{+} $ and $\G_{-} $.
According to (\ref{eq:evenoddsolution}) or (\ref{FEX}), these
equations imply the following boundary conditions on the upper edge of
the cut $[-1,1]$,
\be\label{eq:funcexactongammas}
  \G_{+}  (u+i0)+\G_{-}  (-u+i0)=0,\;\; u\in [-1,1]\, ,
\ee
 or, in terms of the  variable $x$, 
 \be \la{FEGX} \G_{+}  (-1/x)= - \G_{-} 
 (x) \, . 
  \ee
The last equation should hold on the arc $| x|=1 , \Im x >0$.

Hence the solution of the BES equation must be among the solutions of
the functional relation.  Note that this equation is exact in the
sense that in the derivation we did not assumed that $\e$ is small.

Of course this relation has a huge set of solutions.  The physical
solution is distinguished by imposing its analyticity properties in
the vicinity of the singular points $u\to\infty$ and $u=\pm 1$.  The
extra conditions that single out the physical solution are formulated
in terms of the original resolvents $R_\pm = {1\over 2}(R\pm H)$, or
equally the functions $r_\pm(u)$ defined by (\ref{defrpm}).  Namely,
the functions $r_\pm$ must be analytic everywhere outside the cut
$[-1,1]$, where they have square-root singularities, and behave at
infinity according to (\ref{asyRr}).  This conditions determine the
analytic properties of $\G_{\pm} $, which are related to $r_\pm$ by
\be\la{Gmr}
     \G_{-} (u)
     = \hf r_+(u) - \hf r_-(u) + \sum_{n=1}^\infty r_+(u+2i n\e)
      \, ,
     \\
	  \la{Gpr} \G_{+} (u) =2i\e + \hf r_+(u) +\hf r_-(u) +
\sum_{n=1}^\infty r_-(u+ 2i n\e ) \, .  \ee
 Since the functions $r_\pm$ have a square root cut along the interval
 $[-1,1]$ of the physical sheet, all singularities of the functions
 $\G_{\pm} $ are of square root type.

\section{Perturbative solution at strong coupling}

In this section we obtain the perturbative solution for the resolvent. 
First we consider the limit $\e\to 0$ with $u$ fixed. This limit corresponds
 to the plane waves (PW) or giant magnons (GM) regimes, depending 
 on the interval where the rapidity takes its values (Fig. 2).  
 The distribution of Bethe roots in the PW and the GM regimes is given by
 two different analytical expressions, but for the resolvent they they
 are related by analytical continuation.

\vskip 1cm

%%%%%%%%%%Fig.2 %%%%%%%%%%%%%%%%%%%% 
 \begin{center}
\includegraphics[scale=0.5]{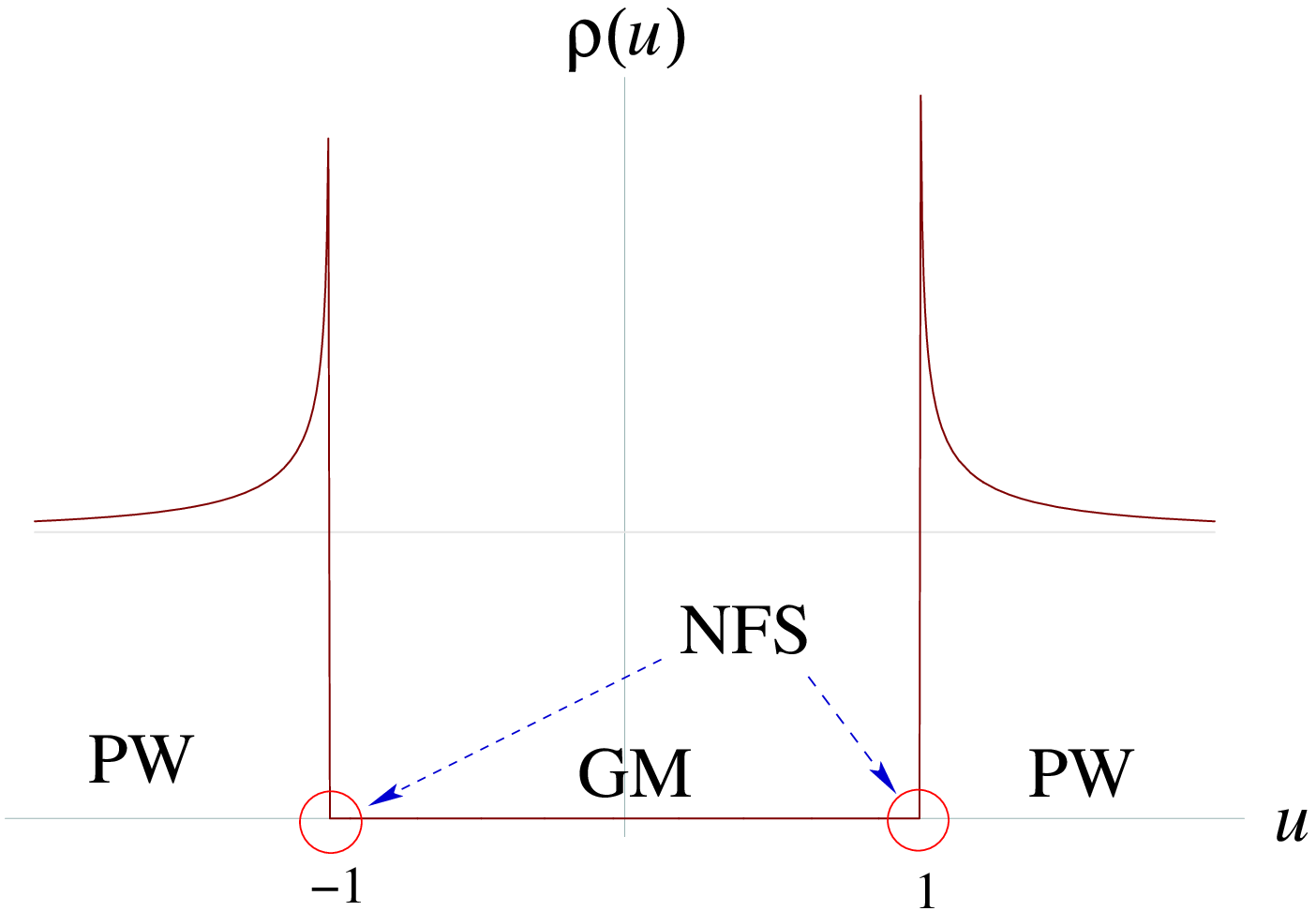} 
\end{center}
\begin{center}  
{\small Fig.  2:  The physical density $\rho(u) = {2\e/
  \pi} - \s(u)$ in the strong  
coupling limit  and the three regimes:
  plane waves (PW) for $u<-1$ and $u>1$,  giant  magnons (GM)
   for $-1<u<1$,  and
  near flat space (NFS) in the vicinity of the points  $u=\pm 1$.  }
\end{center}
%%%%%%%%%%%%%%%%%%%%%%%%%%%%%%%%%   
  \vskip 1cm

 It happens that in the strong coupling limit, 
and in all orders in $\e$, the intricate cut structure of the resolvent  and 
the related functions can be replaced by a single cut $u\in[-1,1]$, but with 
fourth order instead of second order branch points at $u=\pm 1$.   
Furthermore, an important simplification stems from the fact that in the 
PW/GM regime the combinations $\G_\pm(u)$ have definite parity,
\be\la{parityGpm}
\G_{\pm} (-u) = \pm \G_{\pm} (u) \, .  
\ee
This will allow us to write the general solution of the linearized
 BES equation.

Since the equations are homogeneous, the general solution is a linear 
combination of all particular solutions with arbitrary coefficients 
$c_n$, which are functions of the coupling constant.
The behavior of the resolvent at $u\to\infty$ gives one linear constraint on the 
coefficient functions $c_n(\e)$, which is not sufficient to determine them.  

The rest of the information is supplied by the conditions on the analytic
propertirs of the solution in the vicinity of the singular points $u=\pm 1$.
For this purpose we  blow up the vicinity of the the two singular 
points so that the cut structure of the resolvent reappears. Instead of 
keeping $u$ fixed, 
we take the limit $\e\to 0$  either with $z=(u- 1)/\e$ fixed or with  
$\bar z=(u- 1)/\e$ fixed.   This strong coupling limit corresponds to 
the near flat space (NFS) regime \cite{Maldacena:2006rv}.
Then we compare the power series expansion at $z=0$, which follows
from the analytic structure of the exact solution, with the expansion at $z=\infty$,
which is determined by the perturbative solution in the PW/GM regime.
The requirement that the two expansions match with each other is sufficient
to determine both of them, order by order in $\e$.  Technically it is more 
advantageous to compare the inverse Laplace transforms for which the 
shift operator $D$ becomes diagonal.  A recurrence procedure, analogous
 to that of \cite{Basso:2007wd}, allows to obtain analytically the density 
 of Bethe roots in any order in $\e$, both in the PW/GM and NFS regimes.
We check that the result of \cite{Basso:2007wd} for the universal scaling
 function is correctly reproduced.

\subsection{General form of  the  solution in the PW/GM regime}
 
Let us first prove the symmetry property (\ref{parityGpm}).
%This condition restricts considerably the set of the solutions of the
%functional equation (\ref{eq:funcexactongammas}).   
For that we use the expression of $\G_{\pm} $ in terms of the
functions of definite parity $r_\pm(u) =\pm r_\pm(-u) $, given by
(\ref{Gmr}) and (\ref{Gpr}).  We observe that the combinations
\be\la{Gperiodic}
%{\G_{\pm}  (u) \pm   \G_{\pm}  (-u) }  &=&  4i\e +   r_\pm(u)
%\mp \sum_{n\ne 0}  \sgn(n)\, r_\mp( u+ in 2\e),
%\\
\G_{\pm}  (u) \mp  \G_{\pm}  (-u)
        &=&\mp  \sum  _{n\in\MZ} r_\mp( u+ 2in \e)
         \,
\ee
are periodic functions with period $2i\e$.  From here and from the
fact that $r_\pm(u)\sim 1/u^2$ at infinity it follows that the r.h.s.
(\ref{Gperiodic}) vanishes in the limit $\e\rightarrow 0$ up to
non-perturbative terms.  To see that we perform Poisson resummation.
Assuming that $\Re u>1$, we have
 \be
\suml_{n=-\infty}^{\infty}r_\pm (u+2in\e)
%&=&\frac{1}{2\e}\suml_{n=1}^{\infty} \, e^{i\pi n p/\e}
%\frac{dp}{2\pi }\ r_\pm (u+ip)\\ &=&\frac 1\e\suml_{n=-\infty
%}^{\infty}e^{-\pi n u/\e}\ \oint \frac {dv}{2\pi i}\, e^{\pi n v/\e
%}r_\pm(v)\\
&=& \frac 1\e\suml_{n=1}^{\infty}\, e^{-\pi n u/\e} \oint \frac
{dv}{2\pi i}e^{\pi n v/\e }r_\pm(v)\, , \ee
where the integration contour closes around the physical cut $[-1, 1]$
of $r_\pm$.
% We used the fact that $r_\pm(u)\sim 1/u^2$ at infinity, which
% implies that the $n=0$ term vanishes.
The series in $e^{-\pi n u/\e}$ is rapidly convergent when $u>1$ and
diverges at $u=1$.
% since the integral on the r.h.s., which is in fact the inverse
% Laplace transform of $r_\pm$, behaves as $e^{2\pi n/\e} r_n$, where
% $r_n$ has a power-like behavior at large $n$.
When $\Re u<-1$ we get a similar expansion, but the opposite sign in
the exponents.  In both cases the result is exponentially small except
at the points $u=\pm 1$.

Therefore, if we neglect these non-perturbative corrections, the
solution should have the additional symmetry (\ref{parityGpm}).  Then
the functional equation (\ref{eq:funcexactongammas}) can be replaced
by a simpler one,
\be\la{FEGpm}
   \G_{+}  (u+i0)&=&+\G_{-}  (u-i0)\no\\
  \G_{+}  (u-i0)&=&-\G_{-}  (u+i0)\, .
\ee
We remind that these equations are valid on the cut, where
$u\in[-1,1]$.  From here it follows that $\G_{\pm} $ are obtained as
different branches of the same meromorphic function, defined on a
four-sheet Riemann surface.
 
In terms of the global parameter of the Riemann surface, $s $, the
functional equations get the form of periodical conditions
\be\la{FeqGa} \G_{\pm} (s\pm i\pi) &=& \pm \G_{\mp} (s) \, .  \ee
It is convenient to work with the combinations
\be\la{defGGpm} G_{\pm} = \G_{+} \pm i \G_{-} \, , \ee for which
(\ref{parityGpm}) and (\ref{FeqGa}) take the form \be\la{EqsGGpm}
G_{\pm} (-s)=G_{\mp} (s)\, , \quad G_{\pm} (s+ i\pi) = \pm i G_{\pm} (
s)\, .  \ee
We can represent the general solution in the form of the series:
\be
\la{solGs} G_{\pm}  (s)&=& 2i\e \sum_{n\in \MZ} c_{ n} (\e) \, e^{\pm
(2n+1/2)s} \no\\
& =&  2i\e    \sum_{n\in  \MZ}   c_{ n}  (\e) \,
 \({u+1\over u-1}\)^{\pm n  \pm {1\over 4}}
 \, .\ee
\def\sol{(\ref{solGs}) }

We will first obtain the general form of the solution in the  PW/GM regime the branch points condense into continuous lines
starting at the points $u=\pm 1$ and the resolvents are described, as
we will see later, by meromorphic functions with a single pair of
branch points at $u=\pm 1$.

The perturbative solution \sol is valid in the limit where the
distance between the subsequent branch points vanishes and the
infinite sequence of simple branch points starting at $u=\pm 1$
produces a fourth order branch singularity at $u=\pm 1$.

The solution has three singular points, $u=1, u= -1$ and $u\to
\infty$.  As usual in such kind of problems, the coefficients
functions $c_n(\e)$ in the series \sol will be evaluated by matching
with the asymptotic behavior of the solution at the singular points.

  \subsubsection{ The density in the  GM regime}
 
The general form of the solution given by \sol is sufficient to
determine perturbatively the density in the giant magnon regime.
Indeed, inspecting each of the terms, one can verify that the value of
the resolvent on the interval $-1<u<1$, and therefore the density, is
constant and is given by the leading order.

This fact is actually a direct consequence of the equations
(\ref{FeqGa}) and the symmetry (\ref{parityGpm}).  Indeed using the
anti-symmetry of the resolvent $R_{\rm phys}$, we can express the
fluctuation density in terms of the values of the resolvent above the
real axis:
\be \la{densRes}
\s(u)&=&- {1\over 2\pi i}[R_{\textrm {phys}}(u+i0)+R_{\textrm
{phys}}(-u+i0)]\\
&=& - {1\over 2\pi i}[R(u+i\e)+R (-u+ i\e)] \, .  
\ee
Further, by the definition (\ref{defGpm}), the resolvent $R_{\rm
phys}$ is expressed in terms of $\G_{\pm} $ as
  \be R_{\textrm {phys}}=-2i\e+\frac {2}{D+D^{-1}}\( D^{\frac
  12}\G_{-} +D^{-\frac 12}\G_{+} \)\, \qquad (\Im u >0) \no \ee
where $ D $ is the shift operator defined by (\ref{defD}).  Applying
the functional equation (\ref{FEGpm}), we see that all the terms on
the r.h.s. of (\ref{densRes}) except the constant term cancel and
therefore to all orders in $\e$
\be
  \s(u)={2\e}/\pi , \quad  u\in [-1, 1] \,.
\ee
 This means the distribution of Bethe roots has a gap on the interval
 $[-1,1]$.  The physical density (\ref{defrho}), which gives the
 distribution of Bethe roots, vanishes to all orders in $\e$ in the GM
 regime.

\subsubsection{Expansion at $u=\pm 1$ and a scaling condition for the
coefficients}

Let us examine the behavior of the solution \sol near the singular
points $u=\pm 1$.  We mentioned that the strong coupling limit is not
uniform in $u$.  The strong coupling solution have different
properties in the limit considered above,
\be
\la{PWGMlimit}
\e\to 0  \qquad  {\rm with} \ u \ {\rm fixed} \hskip 1cm{\rm (PW/GM)}, 
\ee
and the limit
\be
\la{NFSlimit}
\e\to 0 \qquad {\rm with} \  {u^2-1\over \e}  \ {\rm fixed}\ 
  \hskip 1cm {\rm (NFS)}\, .
\ee
The singular behavior at $u=\pm 1$ in the PW/GM limit is an artifact
of the rescaled rapidity (\ref{newu}).  If we take the NFS limit
(\ref{NFSlimit}), the solution for the density must be integrable at
$u=\pm 1$.  It is obvious that the strong coupling expansions in the
two limits do not match since the solution \sol\ gives non-integer
powers of $\e$ when considered near $u=\pm 1$.

Our analysis of the analytical properties of the solution allows us to
determine its general form near $u=\pm 1$.  The conditions that it
goes smoothly into the solution \sol obtained for the rest of the
complex plane will be used in the next section to fix the coefficients
$c_n$.

The complex variables relevant for the vicinity of the points $u=1$
and $u=-1$ are
 \be z= {u-1\over 2\e}, \qquad \bar z =- {u+1 \over 2\e}\, .  
 \ee
 The variable $z$ coincides, up to a shift by $ 2g$, with the original
 (before rescaling by $2g$) rapidity in the BES equations.

\vskip 1cm

%%%%%%%%%%Fig.3 %%%%%%%%%%%%%%%%%%%% 
 \begin{center}
\includegraphics[scale=0.6]{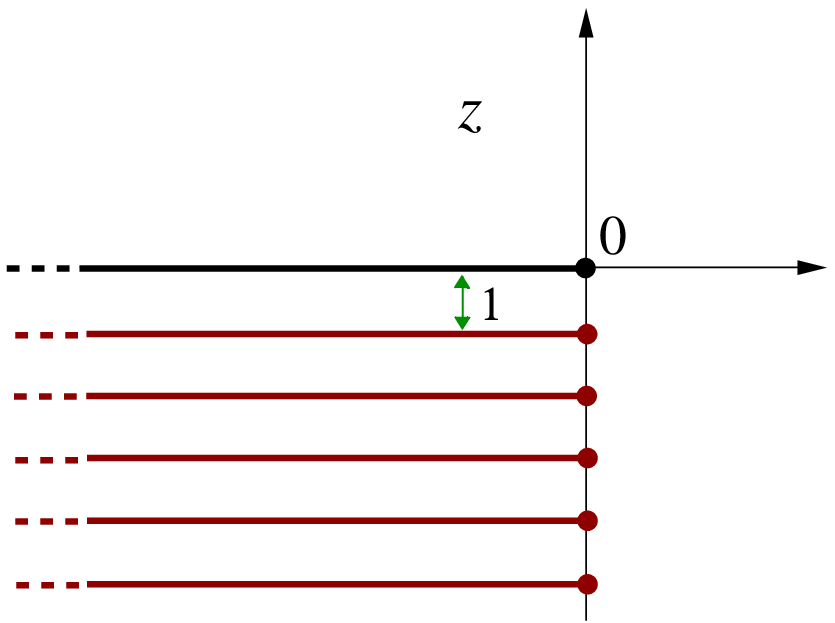} 
\end{center}
\begin{center}  
{\small Fig.  3:  Physical sheet for $R_\pm(u)$ in the NFS limit.  }
\end{center}
%%%%%%%%%%%%%%%%%%%%%%%%%%%%%%%%%   
  \vskip 1cm

In the NFS limit the cuts become semi-infinite, with the branchpoints
placed at at $z= 0, -i, -2i, \dots$, as shown in Fig. 3.
The functions $r_\pm(z)$ have by construction an integrable square
root singularity at $z=0$.  Therefore they can be expanded at small
$z$ as
\be\la{smallz}
r_\pm (z) = \sum _{n\ge 0} b_n^\pm(\e) \, z^{n-1/2} + \sum
_{n\ge 0} d^\pm_n(\e) \, z^{n} \, \, \qquad (\, |z|<1\, ). 
\ee

The compatibility of the expansions (\ref{smallz}) and \sol imposes
strong restrictions on the coefficient functions $c_n(\e)$.  In
particular, each term of the expansion \sol must have a non-singular
limit $\e\to 0$ when expressed in terms of the variable $z$ or $\bar
z$.  This means that the coefficients $c_n(\e)$ must scale as
$\e^{|n|}$, so that their Taylor series have the form
 \be
    \la{DSL}
    c_n (\e) =\e^{|n| }\, { \a_{n}(\e) },
    \qquad
    \a_n(\e) =   \sum_{p=0}^\infty \a_{n,p}\, \e^p
 \qquad (n\in \MZ)\,  .
 \ee
We arrive at the following expression of the general solution in terms
of the shifted rapidity variable $z$:
    \be
 \la{solGr}
  G_{\pm}  (z)
  =  2 i\e  \sum_{n\in\MZ}     \e^{|n|}  \, {\a_{ n} (\e)  }   \,
 \({ 1+\e z\over  \e z}\)^{\pm n  \pm {1\over 4}}\, 
 .
\ee
The strong coupling expansion of $G_\pm$ with $z$ kept fixed is
different than the expansion of the solution with $u$ fixed,
(\ref{solGs}).  In particular, it contains fractional powers of $\e$.
The resolution of this paradox is in the non-uniformity of the strong
coupling expansion with respect to the rapidity variable $u$.  Near
the singular points $u=\pm 1$ the strong coupling expansion should be
performed according to the prescription (\ref{NFSlimit}) and not
(\ref{PWGMlimit}).  The series (\ref{solGr}) should be understood as
an expansion at large $z$, possibly asymptotic, of the true solution,
whose small $z$ expansion is given by (\ref{smallz}).  The
compatibility of (\ref{solGr}) and (\ref{smallz}) is studied more
easily for the inverse Laplace transforms.  This will be done in the
next section where we will see that demanding that the two series are
compatible fixes uniquely the coefficients of both of them.

\subsubsection{ Expansion at $u=\infty$ and universal scaling function}

By construction, the solution \sol expands at infinity as
\be
\la{expGpminf} G_{\pm} (u) = \sum _{n\ge 0} {W^\pm_n\over u^{n }} \ee
Comparing the series with the large
$u$ asymptotics (\ref{largeuR}) of the physical resolvent
\be
  R_{\textrm {phys}}
  %=-2i\e+\frac {2}{D+D^{-1}}\( D^{\frac 12}\G_{-} +D^{-\frac 12}\G_{+} \)
  =-2i\e+\frac {\sqrt{D}}{1+D^{2}}\[(1-i  D  )G_{+} 
  +  (1+i  D )G_{-}   \]\,  ,
\ee
 we fix the first two coefficients $\hf(W^+_0+W^-_0)= 2i\e$ and
 $\hf(W^+_1+W^-_1)= f(g)/16 g^2$.  This yields a constraint for the
 expansion coefficients,
\be\la{sumcn}
 1=  \sum_{n\in  \MZ}   c_{ n}  (\e) 
  \equiv  \sum_{n\in  \MZ}   \e^{|n| }\, { \a_{n}(\e) }     \, ,
  \ee
 and the expression of the universal scaling function $f(g)$ in terms
 of $c_n$:
 \be
 \la{fthrucn}
 f(g) &=& {1\over \e} \sum_{n\in\MZ} (4n+1)c_n
 ={1\over \e} +  {1\over \e} \sum_{n\ne 0} 4nc_n\no \\
& =&
 {1\over \e} +  4 \sum_{n\ne 0}   \e^{|n| -1}\, n\,  \a_{n}(\e) 
 \, .
\ee

\subsubsection{The leading order in the PW/GM limit}

It follows from the scaling (\ref{DSL}) that the solution at the
leading order is given by the $n=0$ term of the series
\be\la{Slead} 
G_{\pm} (s) = 2i \e \, \({u+1\over u-1}\)^{\pm {1\over
4}}\, .  
\ee
The constraint (\ref{sumcn}) gives $c_0(0)= 1$ and the universal
scaling is given by the $n=0$ term in (\ref{fthrucn}):
 \be
 f(g) = {1\over \e} = 4g\, .
 \ee

Written for the resolvent and in terms of the variable $x(u)$, the
leading order solution (\ref{Slead}) is
\be
  R_{\e=0}(u)
 & =& {-2i\e} \( 1-{ 1\over  \sqrt{1-{1/ x^2}}}
 + i{1/x\over \sqrt{1-{1/ x^2}}}   \) \, .
\ee
The density $\s(u)$, related to the resolvent by (\ref{density}),
agrees with the AABEK solution \cite{Alday:2007qf, Kostov:2007kx}.

 \subsection{Inverse Laplace transform of the solution}
 
The relation (\ref{Gperiodic}) involves the shift operator and
therefore looks simpler for the Fourier transformed quantities.
However, in order to be able to exploit the analytic properties of the
general solution we perform instead an inverse Laplace transformation.
Since the functions $g_{\pm}$ and $ G_{\pm} $ are analytic for $\Re
z>0$, we can define the Laplace transformation and its inverse
 \be\la{invlaplace}
   f(z) = \int _0^\infty d\ell \ e^{ - z\ell}\,  \tilde f(\ell)\, \qquad
 \tilde f(\ell) ={1\over 2\pi i}
\int_{i \MR + 0}  d  z \, e^{z\ell}\,  f(z)\,
.
\ee
Similarly we can define the inverse Laplace transformation for the
variable $\bar z$ having as the origin the left branch point.

Introduce, similarly to (\ref{defGGpm}), the linear combinations
\be
g_{\pm} = r_+\mp i r_-\, .
\ee
Then from (\ref{Gmr}) it follows that the functions $g_\pm(z)$ are
related to $G_{\pm} $ to by
  \be\la{ggpmGGpm} g_{\pm} = {1\pm i \over D\mp i}\, (D-1) \, G_{\pm}  \, ,
  \ee
where $D= e^{i\p_z}$ is the shift operator defined in (\ref{defD}).
For the inverse Laplace images $\tilde g_\pm$ and $\tilde G_{\pm} $
this relation takes the form
\be
\tilde g_{\pm}(\ell)= {\sqrt{2} \sin({\ell\over 2})\over \sin({\ell\over
2} \pm {\pi\over 4} )} \, \tilde G_{\pm} (\ell) \, . 
 \la{FEell}
\ee

Our aim is to use the relation (\ref{FEell}) to investigate the
compatibility of the general solution \sol with the expansion
(\ref{smallz}) at $z=0$, which in the $\ell$-space becomes expansion
at $\ell\to\infty$:
\be\la{expellrpm} \tilde g_{\pm} (\ell ) &=& \ell ^{-1/2} \sum _{n\ge 0}
\, \tilde g_n^\pm \, \ell^{-n }+ \sum_{ n\ge0 } \tilde h^\pm_n\,
\ell^{-n-1} 
%\, \qquad (\, \ell \gg 1\, )\, ,\\ \tilde G_{\pm} (\ell)&=& \ell^{
%\pm {1/4}} \sum_{n\in \MZ} { a_n^\pm \over \, \G( n \pm {1\over
%4})}\, \ell ^{ n -1}\, \qquad (\, |\ell | <1\, )
% 
 \, .
  \la{expGl}
 \ee
It follows from the analytic properties of the resolvents in the
rapidity space that, in the NFS limit, $ \tilde g_{\pm}(\ell)$ are
analytic everywhere except for the negative real axis, while $\tilde
G_{\pm} (\ell) $ are analytic everywhere on the positive real axis.
We sketch the proof in Appendix A. The explicit expression for the
inverse Laplace transform of (\ref{solGr}) is a series of confluent
hypergeometric functions of the first kind
\be\la{soll}
&&
\tilde G_\pm(\ell) =   \pm 2
    i     \sum_{n\in \MZ }    \e^{|n|}  \a_{ n}  (\e) \,
    \( n +\qt \)
    \, _1F_1\left(1\mp  \qt  \mp n;2;- {\ell/\epsilon }\right)
 \, . \ee
%[E. E. Kummer (1836)]

The PW/GM corresponds to keeping $\z \equiv \ell/\e$ finite when
$\e\to 0$ while the NFS regime is obtained when keeping $\ell$ fixed.
%The expansion of \sol at $u\to\infty$ translates into the expansion
%of the inverse Laplace image at $\z=0$: \be _1F_1(a,2, -\z)=
%\sum_{k=0}^\infty {(a)_k \over k!(k+1)!} (-\z)^k \ee with $(a)_n =
%a(a+1)\dots (a+n-1)= {\G(a+n)\over \G(a)}$.
In the NFS limit we expand in $\e$ with $\ell $ fixed.  Therefore, in
order to compare with (\ref{expGl}), we are going to use the
asymptotic expansion the limit $\ell/\e \to\infty$, where the solution
has an essential singularity:
 \be\la{asymptF}
 \, _1F_1\(a;2;-{\ell/ \e}\)&\propto& \( {\ell/ \e}\)^{  -a}  
  \, _2F_0\left(a,a-1;;\ {{\e /\ell}}\right)
 \, /\Gamma (2-a)\no\\
   &+& e^{-{\ell/\e}} \( - {\ell/ \e}\) ^{a-2}     \,
   _2F_0\left(1-a,2-a;;\ {{\e/ \ell}}\right)/\Gamma (a)
   \, .
\ee
The asymptotic expansion of the inverse Laplace image of the solution
in this limit is evaluated using by (\ref{asymptF}).  As far as we
interested in the perturbative solution, we can neglect the second
exponentially small term in (\ref{asymptF}) and write
%   %
% \be\la{asymex} \, _1F_1\(a;2;-{\ell/ \e}\)&\propto& \( {\ell/
% \e}\)^{ -a} \, _2F_0\left(a,a-1;;\ {{\e /\ell}}\right) \, /\Gamma
% (2-a)\no\\
%% &=& {(\e/\ell)^a \over \G(2-a)} \sum_{p=0}^\infty {\G(a+p)\G(a+p-1)
%% \over p!  \G(a)\G(a-1)} \({\e/ \ell}\)^p\\
% &=& (\e/\ell)^a \sum_{p=0}^\infty \frac{ (a)_p(a-1)_p }{p!  \,
% \Gamma (2-a)} \({\e\over \ell}\)^p \, \ee with $a = \mp n +1 \mp \qt
% $ and $(a)_p = a(a+1)\dots (a+p-1)= {\G(a+p)\over \G(a)}$.
%
\be\la{lexp}
\tilde G_\pm(\ell)  
%=\pm 2 i \sum _{n\in\MZ}
%\e^{|n|} \a_n(\e) { (\e/\ell)^{1\mp {1\over 4} \mp n} (n+\qt )
%\over \G(1\pm {1\over 4}\pm n)}
% \,
%   _2F_0\left( 1\mp \qt  \mp n,\mp \qt  \mp n ;;\ {{\e\over \ell}}\right)
= 2 i \sum _{n\in\MZ} \e^{|n|} \a_n(\e) { (\e/\ell)^{1\mp {1\over 4}
\mp n} \over \G( \pm {1\over 4}\pm n)} \, _2F_0\left( 1\mp \qt \mp
n,\mp \qt \mp n ;;\ {{\e/ \ell}}\right) \, .  \ee
In the leading order in  $\e$  
\be\la{leadingNFSR} \tilde G_\pm(\ell) =2i (\e/\ell)^{1\mp{1\over
4}}\( \sum_{n= 0}^\infty {\a_{\pm n,0}\over \G( n\pm {1\over 4} )} \,
\ell^{ n } + {\cal O} ( \e)\) \, .  \ee
We see that even in the leading order the resolvents scale in the NFS
regime as fractional powers of $\e$ and are linked to the whole
perturbative series in the PW/GM regime.  In the leading order the sum
on the r.h.s. of (\ref{leadingNFSR}) contains only non-negative powers
of $\ell$, but in the next orders in $\e$ more and more negative
powers of $\ell$ will appear.

Now we represent, as in \cite{Basso:2007wd}, the ratio of the sine
functions in (\ref{FEell}) as
\be { \sin({\ell\over 2})\over \sin({\ell\over 2} \pm {\pi\over 4} )}=
{S_\pm(\ell)\over T_\pm(\ell)}\, , 
\ee
where $S$ and $T$  represent ratios of Gamma functions:
\be S_\pm(\ell) =\pm {\G({1\over 2}+ {\ell \over 2\pi} \mp {1\over
4})\over \G({\ell\over 2\pi} )}\, , \qquad T_\pm(\ell)&=& {\G(1-{\ell
\over 2\pi})\over \G( {1\over 2}-{\ell\over 2\pi} \pm {1\over 4})} \,
.  \ee
 If we rewrite the equation (\ref{FEell}) as
\be\la{FEM} { \tilde G_{\pm} (\ell)\over T_\pm(\ell) } &=&{1 \over
\sqrt{2}} \ { \tilde g_{\pm}(\ell)\over S_\pm(\ell) } \la{Fuec}\, ,
\ee
then the l.h.s. is analytic everywhere except the negative real axis,
while the r.h.s. is analytic everywhere except the positive real axis.
As a consequence, neither of the sides has poles and the only
singularities can be branch points at $\ell=0$ and $\ell=\infty$.
This means, in particular, that the expansion of the r.h.s. as a power
series at $\ell =\infty$ coincides with the expansion of the l.h.s. at
$\ell =0$.

To evaluate the coefficients of the two power series we need to expand
$S_\pm$ at $\ell=+\infty$ and $T_\pm$ at $\ell=0$,
\be S_\pm(\ell) &=& \pm \( \ell /2 \pi \)^{{1\over 2}\mp {1\over 4}}
\( 1+ \sum_{n=1}^\infty S ^\pm_n \ell^{ -n} \) \, , \la{Sexp} \\
  T_\pm (\ell)&=&\frac{1}{\Gamma \left(\frac{1}{2} \pm
  \frac{1}{4}\right)} \(1+\sum_{n=1}^\infty T^\pm_n \ell^n\) \, .
  \la{Texp} 
  \ee
As it should, the series expansion of the l.h.s. of (\ref{FEM}) at
$\ell=0$ contains exactly the same fractional powers as that for the
r.h.s. at $\ell=\infty$.  To get rid of these fractional powers, we
multiply both sides of (\ref{FEM}) by $(\ell/\e)^{1\mp 1/4}$ and write
 \be
 \la{seriesGT} {\tilde G_{\pm} (\ell)\over\tilde T_\pm(\ell)} \(
 {\ell / \e}\)^{ 1 \mp {1\over 4}}\ = \sum_{n\in \MZ} \ C_{n}^\pm
 (\e)\ \ell^{-n }\, , 
 \ee
where the coefficients $C_n^\pm(\e)$ should be understood as formal
series in $\e$,
\be\la{TaylC}
C_n^\pm(\e)= \sum_{p=0}^\infty C_{n,p}^\pm \, \e^p\, .
\ee
 From (\ref{expellrpm}), (\ref{Sexp}) and the relation (\ref{FEM}) we
 deduce that the coefficients in front of the non-negative powers of
 $\ell$ vanish,
   \be \la{Constraint}
   C_n^\pm (\e) = 0  \quad  {\rm for} \quad n=-1,-2, \dots\, \ .
   \ee   
Solving these contraints (\ref{sumcn}) and (\ref{Constraint}) order by
order in $\e$ one can evaluate recursively the Taylor coefficients
$\a^\pm_n$ of the series (\ref{DSL}).  The recurrence procedure is
possible because at each order in $\e$ the sum on the r.h.s. of
(\ref{seriesGT}) contains only a finite number of negative powers of
$\ell$.

We have learned that the general solution of the BES equation in the
NFS limit is of the form
\be\la{exNFS} \tilde G_\pm(\ell)&=& \({\ell/\e}\)^{-1\pm {1\over 4} }
T_\pm(\ell) \ \sum_{n\ge 0}C_{n}^\pm (\e)\ \ell^{-n }\, , \\
 \tilde g_{\pm}(\ell) &=& \sqrt{2}\, \({\ell/\e}\)^{-1\pm {1\over 4} }
 \, S_\pm(\ell ) \sum_{n\ge 0} \ C_{n}^\pm (\e)\ \ell^{-n }\, , \ee
with computable coefficient functions given by the formal Taylor
series (\ref{TaylC}).  Comparing the expansions of (\ref{lexp}) and
(\ref{exNFS}) at each order in $\e$ and imposing the condition
(\ref{sumcn}) we can evaluate both sets of coefficients $\a_{n,p}$ and
$C^{\pm}_{n,p}$.  We show below how the procedure works for the
leading order.

\subsubsection{The leading order in the NFS limit}

In the leading order in $\e$ the series expansion of $\tilde G_\pm$ at
$\ell=0$, given by (\ref{leadingNFSR}), contains only non-negative
powers in $\ell$.  Therefore the sum on the r.h.s. of (\ref{seriesGT})
contains only the term with $n=0$, and we have
\be
\la{leadingNFSRbis} 
(\ell/\e)^{1\mp{1\over 4}} \tilde G_\pm(\ell)
=2i \sum_{n= 0}^\infty {\a_{\pm n,0}\over \G( n\pm {1\over 4} )} \,
\ell^{ n } \, = T_\pm(\ell) \, C_{0,0}^\pm \, .  
\ee
From the constraint (\ref{sumcn}), which in the leading order gives
$\a_{0,0}=1$, we evaluate
  \be C^+_{0,0}= {2i} {\G({3\over 4})\over \G({1\over 4})}\, , \ \
  C^-_{0,0}= {2i } {\G({1\over 4})\over \G(-{1\over 4})} \, .  \ee
For  the other coefficients  we find 
\be 
\a_{\pm n,0} &=& {\G(n \pm {1\over 4})\over \G(\pm {1\over 4})} \
T^\pm _n \, , 
\ee
 where $T^\pm _n$ are the coefficients in the expansion (\ref{Texp}),
  \be
  T^+_{1} &=&  \frac{\pi -6 \log 2}{4 \pi } 
  \, , \no\\
  T^-_1 &=&  -\frac{\pi +6\log 2  }{4 \pi   }\, ,
  \no\\
T^+_{2}&=&
   \frac{ 
    96 K-7 \pi ^2-36 \pi \log 2+108 (\log
2)^2
}{96  \pi ^2} \, ,\no\\
T^-_{2}&=&    \frac{96 K+7 \pi ^2-36 \pi  \log 2-108 (\log 2)^2 }{64  \pi ^2}\,  ,
\\
&&{ \dots }  
    \ee

  \subsubsection{The universal scaling function}

 There is no difficulty to carry out  the procedure for  the  higher orders in $\e$. 
 The only diffference will be that  the expansion (\ref{leadingNFSR}) and therefore
 the r.h.s. of (\ref{leadingNFSRbis})   will contain  some  negative powers of $\ell$.  
 We do not go into details because the procedure is technically identical as the 
 one formulated in \cite{Basso:2007wd}. The lowest  orders for $\a_{n}(\e)$ are:
\be
\a_0(\e)& =& 
1-{1\over 8} \e +...\,  ,\\
\a_{1}(\e)&=& \frac{\pi -6 \log 2}{16 \pi }+\frac{-96 K+(7 \pi -12
\log 2) (\pi +6 \log 2)}{128 \pi ^2}\; \e +\dots\,  ,\\\
\a_{-1}(\e)&=& \frac{\pi +6 \log (2)}{16 \pi }
+\frac{-96 K-7 \pi ^2+54 \pi  \log 2+216 (\log 2)^2 }{384 \pi ^2}\; \e+\dots\,  ,\\\
\a_{2}(\e)&=&   \frac{ 5\left(96 K-7 \pi ^2-36 \pi \log 2+108 (\log
2)^2\right)}{ 1536   \pi ^2} +\dots \,  ,\\\
\a_{-2}(\e)&=&\frac{96 K+7 \pi ^2-36 \pi  \log 2-108 (\log 2)^2}{512 \pi ^2}
+\dots\   .\\
\ee
From here we reproduce the result  of \cite{Basso:2007wd}
for the  universal scaling function,
\be
f(\e)&=&
{1\over\e}     + 4   \sum_{n=1}^\infty \e^{|n|-1} n\,  \a_n(\e) \\
%&=&  {1\over\e} \a_{0,0} +  4(\a_{0,1} + \a_{1,0}- \a_{-1,0})+\dots  \\
%&=&  {1\over\e} + 3\a_{1,0}-5 \a_{-1,0})+\dots
%\\
&=& {1\over\e}  -\frac{3\log 2}{\pi }  -\frac{K}{\pi ^2} \e + \dots
\ . \ee

\section{Conclusion}

We have reformulated the Beisert, Eden and Staudacher equation in terms of 
a functional equation obeyed by the resolvent. A similar approach was attempted, although not fully 
exploited, by Kotikov and Lipatov \cite{Kotikov:2006ts}. As shown recently by Basso, 
Korchemsky and Kota\'nski \cite{Basso:2007wd}, in the strong coupling perturbative regime
it is possible to find the general solution as a linear combination of a set of particular functions. 
This is possible because, in the absence of non-perturbative terms, the resolvent possesses 
extra symmetries.
We have shown that the ``quantization condition" of   \cite{Basso:2007wd}, which allows to fix 
the coefficients of the linear combination order by order in the inverse coupling constant $\e$ can be understood as an analyticity condition on the resolvent. The condition that the resolvent
has an integrable singularity of the square root type at the points $u=\pm 1$, together with the 
conditions on the behavior at infinity of the resolvent are sufficient to fix the solution recursively
order by order in $\e$.

Although we have not explicitly investigated the non-perturbative correction, their source
is clearly identified at the level of the resolvent. This object possesses a 
sequence of self-repeating cuts situated at a distance $2\e$ of one another. When $\e\to 0$,
the cuts condense and we are left with a single cut plus a non-perturbative term. 
We leave the investigation of the functional BES equation for a future work.

One of the points of technical importance in the work of  \cite{Basso:2007wd} and in our work 
was to transform the  BES equation into a set of two equations with the so-called undressed kernel appearing linearly. In order to perform this transformation we are led to introduce an auxiliary
density.  It would be interesting to know whether   such a linearization is possible  for the general   
Bethe ansatz equations for ${\cal N}=4$ SYM, and if the auxiliary density  can be given a  physical meaning. Suggestions about the possibility that  the dressed kernel originates from the elimination of an auxiliary set of  Bethe roots have been  made in \cite{Rej:2007vm, Sakai:2007rk}.

 It would be interesting to check if the same  method  can be applied for the integral equations corresponding to other sectors of the ${\cal N}=4$ gauge theory, as well as for the other limits in the $sl(2)$ sector. In particular, it would be interesting to try to reproduce the new universal scaling
function predicted in \cite{Alday:2007mf} and \cite{FRS}
and computed by Roiban and Tseytlin \cite{RT} in the 
so-called slow long string limit.

%%%%%%%%%%%%%%%%%%%%%%%%%%%%%%% 

\bigskip\bigskip

\leftline{\bf Acknowledgments}

\noindent We thank B. Basso and G. Korchemsky for useful discussions.
This work has been partially supported by the European Union through
ENRAGE network (contract MRTN-CT-2004-005616), the by ANR programs
GIMP (contract ANR-05-BLAN-0029-01) and by INT-AdS/CFT (contract
ANR36ADSCSTZ). I.K and D.S. beneficed from the the ``Integrability,
Gauge Fields and Strings'' focused research group at the Banff
International Research Station.
 %%%%%%%%%%%%%%%%%%%%%%%%%%%% 

\appendix

\section{Analytical structure of $\tilde g_\pm$ and $\tilde G_\pm$}

The inverse Laplace transform $\tilde f(\ell)$, originally defined by
(\ref{invlaplace}) for $\ell>0$, can be analytically continued for
complex values of $\ell$ by rotating the integration contour so that
asymtotically $\Re(z \ell)=0$ at large $z$.

\vskip 1cm

%%%%%%%%%%Fig.4 %%%%%%%%%%%%%%%%%%%% 
 \begin{center}
\includegraphics[scale=0.8]{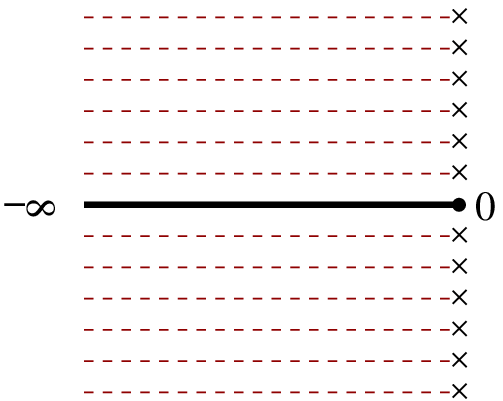} 
\hskip 4cm \includegraphics[scale=0.8]{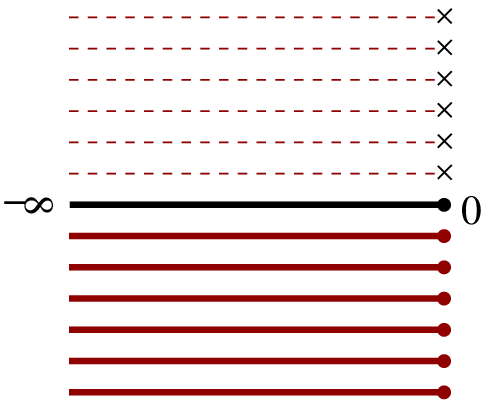}
\end{center}
\begin{center}  
{\small Fig.  4:  Left: Riemann surface for $g_\pm(z)$.
Right: Riemann surface for $G_{\pm} (z)$.  Dots denote branch points on
the physical sheet, crosses denote the positions of the branch points
on the lower sheets.    }
\end{center}
%%%%%%%%%%%%%%%%%%%%%%%%%%%%%%%%%   
  \vskip 1cm

%%%%%%%%%%Fig.5 %%%%%%%%%%%%%%%%%%%% 
 \begin{center}
\includegraphics[scale=0.8]{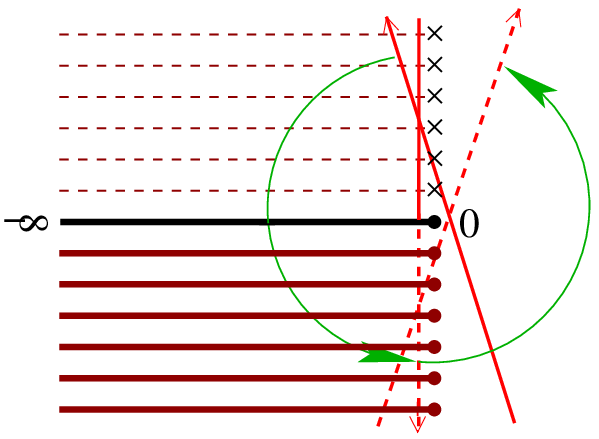}
\end{center}
\begin{center}  
{\small  Fig. 5: Deformation of the integration
contour for the inverse Laplace transform of $G_{\pm} $.   }
\end{center}
%%%%%%%%%%%%%%%%%%%%%%%%%%%%%%%%%   
  \vskip 1cm

 The functions $g_\pm(z)$ have one second-order branch point at $z=0$
 on the physical sheet and an infinite sequence of equidistant branch
 points $z \in i \MZ$ on the second sheet (Fig. 4, left).
 We assume that we are in the NFS limit in which the left endpoints of
 the branch cuts are sent to $-\infty$.  We analytically continue
 $\tilde g_{\pm}(\ell)$ beginning from $\ell$ real positive by
 changing the phase of $\ell$ clockwise.  The contour of integration
 in the definition of the inverse Laplace transform will
 correspondingly rotate counterclockwise.  For $ \ell\in- i \MR_+ $,
 the integration contour will lie along the real axis and above the
 cut of $g_\pm(z)$.  We can further decrease the phase of $\ell$ by
 rotationg the contour so that half of it passes in the lower sheets
 of the Riemann surface.  The procedure can be continued without
 encountering any singularity until the phase of $\ell$ is rotated by
 $\pi$
 \be
 \ell \to e^{i\pi}\ell=-\ell\;.
 \ee
 At this point, the integration contour goes again on the imaginary
 axis, with the lower half now approaching a sequence of branch points
 on the lower sheets.  Since the contour cannot be moved further, we
 deduce that $\tilde g_\pm(\ell)$ have singularities on the negative
 real axis.

The functions $G_{\pm}(z)$ have a sequence of branch points on the
negative imaginary axis on the physical sheet (Fig. 4,
right).  Since $G_{\pm} $ do not decrease sufficiently fast at infinity,
the inverse Laplace transform does not exist for $\ell>0$.  However if
we rotate slightly the contour counterclockwise, as is shown in
Fig. 5, the integral (\ref{invlaplace}) starts to
converge.  In particular, it is well defined on the negative real
axis, when $\ell\to e^{i\pi}\ell$.  After rotating the contour by
angle $\pi$, half of it passes on the second sheet, where there are no
branch points below the real axis.  Therefore we can continue rotating
the integration contour until $\ell \to \e^{2i\pi}\ell$, when we
encounter the branch points on the second sheet, which are on the
positive imaginary axis.  The inverse Laplace transform $\tilde
G_{\pm}(\ell)$ is therefore well defined for $\ell<0$ and the
singularities only occur for $\ell$ on the positive real axis.

\section{Relation with  the  BKK  conventions}

%The functional equation (\ref{FEGpm}) can be written in terms of
% the  variable $x$ as:
%%
%\be
%   \G_{+} (x)&=&\G_{-} (1/x)\;,\; \no\\
%  \G_{-} (x)&=&-\G_{+} (1/x)\; .
%\ee
%%
%The general solution of this equation can be also written in the form
% %
%  \be\label{eq:generalformofperturbativesolution} \G_{+} (u,\e)&=& \frac
%  \e{\sqrt{1-\frac 1{x^2}}}\sum_{p=0}^{\infty}\(\frac
%  {c_{p}^+(\e)\e^{p}}{(1-x^2)^{2p}}+\frac
%  {c_{p}^-(\e)\e^{p}}{(1-x^2)^{2p-1}}\)\no\\
%  \G_{-} (u,\e)&=&\frac \e{ix\sqrt{1-\frac
%  1{x^2}}}\sum_{p=0}^{\infty}\(\frac {c_{p}^+(\e)\e^{p}}{(1-\frac
%  1{x^2})^{2p}}+\frac {c_{p}^-(\e)\e^{p}}{(1-\frac 1{x^2})^{2p-1}}\)
%  \ee 
%%
%which is nothing but inverse Fourier transform (on the half-line) of
%the BKK ansatz (8) in \cite{Basso:2007wd}.  Further, all the arguments
%concerning the NFS limit we used above are valid for this Ansatz as
%well, and give a natural explanation of their quantization conditions.

The functions used in the present paper 
and those used in the paper by Basso,  Korchemsky and  Kota\'nski 
 \cite {Basso:2007wd} are related as follows:
\be
  X_{\rm here}(u)=-i \intoinf dt e^{itu}Y_{\rm BKK}(t)\, , \no
\ee
with
\be
{\begin{tabular}{|c|c|}
  \hline
 %  after \\: \hline or \cline{col1-col2} \cline{col3-col4} ... 
  $X_{\rm here}(u) $&$  Y_{\rm BKK}(t)$\\
     \hline
  $r_+(u)$ & $\g_-(t)$ \\
  $r_-(u)$ & $\g_+(t)$ \\
       \hline
  $\G_+(u) - 2i\e$ & $\hf \G_-(t)$ \\
  $\G_-(u)$ & $-\hf\G_+(t)$ \\
     \hline
\end{tabular}\;
}
\no
\ee

 \end{document}